\documentstyle[ fleqn,epsfig]{article}

\newcommand{\AmS}{{\protect\the\textfont2
  A\kern-.1667em\lower.5ex\hbox{M}\kern-.125emS}}
\begin{document}
\title{
\rightline{\small UL--NTZ 23/96}
\vspace{-10pt}
\rightline{\small  HUB--IEP--96/18} 
\vspace{-10pt}
\rightline{\small DESY 96-113}
$3$--Dimensional Lattice Studies of the Electroweak Phase
  Transition \\
  at $M_{Higgs} \approx 70$ GeV}
\author{M.~G\"urtler$^1$\thanks{guertler@tph204.physik.uni-leipzig.de}, 
        E.-M.~Ilgenfritz$^2$\thanks{ilgenfri@pha2.physik.hu-berlin.de},
        J.~Kripfganz$^3$\thanks{j.kripfganz@thphys.uni-heidelberg.de}, \\
        H.~Perlt$^1$\thanks{perlt@tph204.physik.uni-leipzig.de}
        and
        A.~Schiller$^1$\thanks{schiller@tph200.physik.uni-leipzig.de} \\
{\it $^1$ Institut f\"ur Theoretische Physik, Universit\"at Leipzig, Germany} \\
{\it $^2$ Institut f\"ur Physik, Humboldt-Universit\"at zu Berlin, Germany}\\
{\it $^3$ Institut f\"ur Theoretische Physik, Universit\"at Heidelberg, Germany}
        }

\date{May 31, 1996}
\maketitle
\begin{abstract}
  We study the electroweak phase transition by lattice simulations of
  an effective $3$--dimensional theory, for a Higgs mass of about $70$
  GeV.  Exploiting, among others, a variant of the equal weight
  criterion of phase equilibrium, we obtain transition temperature,
  latent heat and surface tension, and compare with $M_H \approx 35$
  GeV. In the broken phase masses and Higgs condensates are compared
  to perturbation theory. For the symmetric phase, bound state masses
  and the static force are determined.
\vspace*{1cm}
\end{abstract}

\section{Introduction}

Recent lattice studies of the electroweak phase transition
\cite{BunkEA}-\cite{buckowtalk} had been triggered by the interest in
understanding the baryon number asymmetry of the universe. Significant
cosmological consequences would require a sufficiently strong first
order transition \cite{KirzhnitsLi,BuchmuellerEA}. The transition has
to be strong enough, both in order to accomplish a sufficient rate of
baryon generation during the transition and to prevent the wash--out
of baryon number after it is completed. The present quantitative
understanding of possible mechanisms as well as experimental lower
bounds for the Higgs mass make this unlikely within the minimal
standard model. Extensions, in particular supersymmetric ones may
still be viable, however \cite{Cline}-\cite{Laine2}.

A second reason for lattice investigations was the wish to control the
behaviour of perturbative calculations of the effective action. This
quantity is the appropriate tool of (non--lattice) thermal quantum
field theory for dealing with symmetry breaking. Infrared problems
prevent a perturbative evaluation of the free energy in the symmetric
phase to higher loop order. However, the true, non--perturbative
nature of the symmetric phase will be characterized by massive $W$--
and Higgs bound states instead of massless $W$ gauge bosons. A
self--consistent approach to provide masses across the transition, {\it e.g.}
by gap equations \cite{gap}, can improve the ability to calculate
perturbatively the symmetric phase. Gauge field condensates are
another property of the symmetric phase, expected to lower its free
energy density.

An independent method is needed to characterize the electroweak phase
transition even within a pure $SU(2)$ gauge--Higgs version of the
theory.  This model has become a test--field to control the validity of
perturbative predictions over a broad range of Higgs masses. At
present, one is interested to see whether the first order transition
ends somewhere around a Higgs mass $M_H \approx 100$ GeV \cite{karsch,
 Kajantiemay96}. Lattice simulations are not only able to describe
both phases starting from first principles but, moreover, make it
possible to put both phases into coexistence near the phase equilibrium. 
Thus one is able to measure directly quantities like latent heat, surface 
tension, condensates etc. quantifying the strength of the transition.

One approach to lattice calculations of the electroweak transition is
based on an effective $3$--dimensional Higgs model. It is attractive
phenomenologically because it circumvents the problem of putting
chiral fermions on the lattice. Due to dimensional reduction, fermions
as well as non--static bosonic modes contribute to the effective
action. In contrast to QCD, dimensional reduction should work for the
electroweak theory around and above the transition temperature because
$g^2$ is small. For the electroweak phase transition this approach has
been pioneered by Farakos et al. (see {\it e.g.}
\cite{FarakosEA,Kajantieaug95}). This program aims at exploring the
accuracy of dimensional reduction at various Higgs masses by comparing
various parameters of the transition with those of $4$--dimensional
lattice and perturbative approaches. Perturbation theory is necessary
to relate the $4$--dimensional continuum theory to the parameters of
the dimensionally reduced theory and, finally, to the bare coupling
parameters of the lattice action. Dimensionally reduced versions
retain the remnant of the temporal gauge field $A_0$ (as an adjoint
Higgs field) or not (as in this work).

In its simplest version the dimensionally reduced effective theory is
again a SU(2)--Higgs theory with just one doublet. This model
represents whole classes of $4d$ theories but might not be sufficient
to describe all $4d$ variants generically. So far, only this simple
effective theory has been studied by lattice techniques, for several
Higgs self--couplings $\lambda _3$ in units of the $3d$ gauge couplings
squared $g_3^2$ \cite{FarakosEA2, Kajantieoct95, physlett,
  buckowtalk}. Here we extend our analysis \cite{physlett} to a higher
coupling value
\begin{equation}
  \frac{\lambda_3}{g_3^2}=\frac18 (\frac{M_H^*}{80\; \mbox{GeV}})^2
  \label{MH*}
\end{equation}
namely $\lambda _3/g_3^2 \approx 0.095703$, referred to as $M_H^*=70$
GeV, and compare with previous results at this smaller coupling
$\lambda _3/g_3^2 \approx 0.023926$ ($M_H^*=35$ GeV). As expected the
first order nature has become weaker but is still evident.

\section{Relation of the dimensionally reduced model to the $SU(2)$ Higgs
  theory in four dimensions}

The procedure of dimensional reduction in gauge theories consists of
two steps. In a first step, nonstatic modes of the gauge field are
integrated while the temporal component $A_0$ is retained as adjoint
Higgs field. The action is further simplified by integrating over
$A_0$ in a second step. The mass of the latter field is given by the
Debye mass $m_D$. The reduced Higgs theory has the following action
resembling the form of the $4$--dimensional theory
\begin{equation}
  S = \int d^3 x \Big( {\frac{1 }{4}} F_{\alpha \beta}^b F_{\alpha
    \beta}^b + (D_{\alpha} \phi)^+ (D_{\alpha} \phi) + m_3^2 \phi^+
  \phi + \lambda_3 (\phi^+ \phi)^2 \Big),
  \label{eq:cont_action}
\end{equation}
where $\alpha, \beta, b = 1,2,3$. The $3$--dimensional mass parameter
$m_3(\mu_3)$, renormalized at scale $\mu_3$ and the renormalization
group invariant couplings $\lambda_3$ and $g_3$ can be expressed by
dimensional mapping in terms of the running parameters of the
$4$--dimensional theory at the scale $\mu$ and temperature $T$ in the
$\overline{MS}$ scheme. The renormalization scale $\mu$ is most
conveniently chosen as $\mu = \mu_T = 7.05 T$ which puts logarithmic
contributions of bosonic origin to zero.

To make this paper self--contained we collect some formulae for later
use.  As the result of the second step of reduction mentioned above,
the couplings $g_3$ and $\lambda_3$ and the $3$--dimensional mass
$m_3$ can be expressed in terms of the corresponding couplings and
mass of the intermediate theory with $A_0$ included, $\overline{g}_3$,
$\overline{\lambda}_3$ and $\overline{m}_3$ (and the Debye mass
$m_D=\sqrt{5/6}gT$). The latter parameters in their turn can be
written in terms of the temperature and the parameters of the
$4$--dimensional theory:
\begin{eqnarray}
\label{eq:dim_reduct}
g_3^2 & =& \overline{g}_3^2 (1 - {\overline{g}_3^2 \over 24 \pi m_D} ) ,
                                                                  \nonumber \\
\overline{g}_3^2 &=& g^2(\mu_T) \ T \ \big(1 + {g^2 \over 16 \pi^2}
  {2 \over 3}\big) ,
                                                                   \nonumber \\
\lambda_3 &=& \overline{\lambda}_3 -  { 3 \overline{g}_3^4 \over  128 \pi m_D},
                                                                   \nonumber \\
\overline{\lambda}_3 &=& \lambda(\mu_T) \ T  \ \big(1 + {g^2 \over 16 \pi^2}
{3 M_W^2 \over M_H^2} \big) ,
                                                                    \\
m_3^2(\mu_3) &=& \overline{m}_3^2(\mu_3) - {3 \over 16 \pi} \overline{g}_3^2 m_D
+ {\overline{g}_3^4 \over 16 \pi^2}
\big( {15 \over 8} \log {\mu_3 \over 2 m_D} + {9 \over 16}\big) ,
                                                                    \nonumber \\
\overline{m}_3^2(\mu_3) &=& - \nu^2(\mu_T) + T ( {1 \over 2}
 \overline{\lambda}_3
+ { 3 \over 16} \overline{g}_3^2)
+ {T^2 \over 16 \pi^2} \big( {137  \over 96}g^4 + { 3 \over 4} \lambda g^2
\big)
                                                                  \nonumber \\
  &+& {1 \over 16 \pi^2} \big( {81 \over 16} \overline{g}_3^4
  + 9 \overline{\lambda}_3 \overline{g}_3^2 - 12 \overline{\lambda}_3^2 \big)
(\log {3 T \over \mu_3} - 0.348725) .
                                                        \nonumber
\end{eqnarray}
The $4$--dimensional renormalized quantities $g^2(\mu)$ , $\nu^2(\mu)$ and $%
\lambda(\mu)$ have the generic form with one--loop corrections formally
indicated
\begin{eqnarray}
g^2(\mu_T) &=& g_0^2 (1 + \delta g^2)
                                               \nonumber \\
\lambda(\mu_T) &=& { g_0^2 \over 8} {M_H^2 \over M_W^2}
( 1 +\delta \lambda)
                                      \nonumber           \\
\nu^2(\mu_T) &=& {M_H^2 \over 2} ( 1 +  {\delta \nu^2})
\label{eq:4dcouplings}
\end{eqnarray}
with $g_0^2 = 4 \sqrt 2 G_F M_W^2 $. $G_F = 1.16639 \times 10^{-5}
\mbox GeV^{-2}$ is Fermi's constant. The corrections indicated in these
formulae, which depend on $g^2, \mu_T$ and the Higgs and $W$ mass squared
$M_H^2$ and $M_W^2$, can be found in Ref. \cite{Kajantieaug95}.  For
definiteness, we use as the $3$--dimensional renormalization scale
$\mu_3 = g_3^2$ and take $M_W=M_Z=80.6 $ GeV. For the $4$--dimensional 
coupling in the loop corrections we choose $g^2 = g_0^2$.

At $M_H^*=70$ GeV the corrections in 
(\ref{eq:4dcouplings}) 
without fermions
are numerically obtained as follows
\begin{eqnarray}
\delta g^2 & \approx & -0.01343 - 0.01945\log (\mu_T^2/M_W^2), \nonumber \\
\delta \lambda & \approx & -0.01625 + 0.008491\log (\mu_T^2/M_W^2), \nonumber \\
\delta \nu^2 & \approx & -0.01892 - 0.004721\log (\mu_T^2/M_W^2).
\label{eq:4dim_70}
\end{eqnarray}
Similarly, we find for $M_H^*=35$ GeV
\begin{eqnarray}
\delta g^2 & \approx & -0.01281 - 0.01945\log (\mu_T^2/M_W^2), \nonumber \\
\delta \lambda & \approx & -0.05396 + 0.06131\log (\mu_T^2/M_W^2), \nonumber \\
\delta \nu^2 & \approx & -0.07286 - 0.005766\log (\mu_T^2/M_W^2).
\label{eq:4dim_35}
\end{eqnarray}

\section{The $3$--dimensional lattice model}

\subsection{Lattice action}

On the lattice, we study the $SU(2)$--Higgs system with one complex
Higgs doublet of variable modulus. The gauge field is represented by
unitary $2 \times 2$ link matrices $U_{x,\alpha}$ and the Higgs fields
are written as $\Phi_x = \rho_x V_x$. $\rho_x^2= {\frac{1 }{2}}
Tr(\Phi_x^+\Phi_x)$ is the Higgs modulus squared, $V_x$ an element of
the group $SU(2)$, $U_p$ denotes the $SU(2)$ plaquette matrix. The
lattice action is
\begin{eqnarray}
S  &=& \beta_G \sum_p \big(1 - {1 \over 2} Tr U_p \big) -   \beta_H \sum_l
       {1\over 2} Tr (\Phi_x^+ U_{x, \alpha} \Phi_{x + \alpha})
 \nonumber \\
  & & + \sum_x  \big( \rho_x^2 + \beta_R (\rho_x^2-1)^2 \big)
\label{eq:latt_action}
\end{eqnarray}
(summed over plaquettes $p$, links $l$ and sites $x$), with
\begin{equation}
\beta_G = {\frac{4 }{a g_3^2 }}.
\end{equation}
The lattice Higgs self--coupling is
\begin{equation}
\label{eq:betar} \beta_R= {\frac{\lambda_3 }{g_3^2}} \ {\frac{\beta_H^2 }{%
\beta_G}}
\end{equation}
and the hopping parameter can be expressed in the form
\begin{equation}
\beta_H=\frac{2 (1-2\beta_R)}{6+a^2 m_3^2}.
\label{eq:betah}
\end{equation}

The relation between the $3$--dimensional bare mass squared $m_3^2$, expressed
via the lattice couplings (see (\ref{eq:betah})) and the renormalized continuum
mass squared has been worked out by Laine \cite{laine} to two loops
\begin{eqnarray}
\label{eq:laine}
m_3^2(\mu_3) &=& m_3^2 + m_1^2 + m_2^2
                                                  \nonumber \\
m_1^2 &=& {\Sigma \over 4 \pi a} ( {3 \over 2} g_3^2 + 6 \lambda_3) \ , \ \ \
\Sigma = 3.175911
                                                        \\
m_2^2 &=& {1 \over 16 \pi^2} (( {51 \over 16} g_3^4 + 9 \lambda_3 g_3^2 -
12 \lambda_3^2) (\log {6 \over a \mu_3} + 0.09)
                                        \nonumber      \\
 &  & + 5.0 g_3^4 +
5.2 \lambda_3 g_3^2).
                                           \nonumber
\end{eqnarray}
This completes the relation of the lattice parameters of the $3d$
model to the $4d$ continuum parameters.

The lattice model defined by (\ref{eq:latt_action}) is numerically
studied at given couplings $\beta_G, \beta_H$ and $\lambda_3/g_3^2$.
We have used the same simulation algorithms as in our previous
investigations. We have combined a $3$--dimensional Gaussian heat bath
update for the gauge fields with a $4$--dimensional Gaussian heat bath
method for the Higgs field. In the last one, the Gaussian step was
improved for later acceptance by taking the non--linearity into account.
To reduce the autocorrelations near to the phase transition, a heat
bath step as described was followed by several reflections (eight in
practice) for the Higgs (and partly one reflection for the gauge
field).

In the search for the phase transition, bulk variables like the
space--averaged square of the Higgs modulus and the space--average link
\begin{eqnarray}
\label{eq:order_param}
{\rho^2} & = & { 1 \over {L_x L_y L_z}}
\sum_x \rho_x^2  , \nonumber \\
{E_{link}} & = & { 1 \over {3 L_x L_y L_z}}
\sum_{x,\alpha} {1\over 2} Tr (\Phi_x^+ U_{x, \alpha} \Phi_{x + \alpha}) ,
\end{eqnarray}
are heavily used (the average quartic Higgs modulus ${\rho^4}$ is defined 
analogously). Others like the average plaquette
\begin{eqnarray}
\label{eq:aver_plaq} P = {\frac{ 1 }{{3 L_x
L_y L_z}}} \sum_p {\frac{1 }{2}} Tr U_p
\end{eqnarray}
are less useful for this purpose, although they also show a two--state signal
on large enough lattices. In principle, the histograms and discontinuities
of all bulk variables have to be known in order to estimate the strength of
the phase transition, but ${\rho^2}$ proves to be the most important one. It 
has turned out that, due to the confinement property of the symmetric phase, 
histograms of medium size rectangular Wilson loops show a two--state signal 
as well.

While the bulk quantities discussed are measured after each Monte Carlo
iteration, other quantities like Wilson loops and correlation functions are
measured after every 10th iteration.

To reduce the noise in the evaluation of Wilson loops the original
links are replaced by their mean field values \cite{Parisi}. They are 
calculated from the links interacting with the original links through the 
staples. In the case of $SU(2)$ this value is calculated analytically.
This procedure is not applied to the links in the corners of the
Wilson loops.

The Higgs and vector boson masses are measured from correlation functions of
extended operators of length n ($\tau^b$ are the Pauli matrices) 
\begin{eqnarray}
  0^+&: & S_{x,\alpha} (n) =\frac12 Tr (\Phi^+_x U_{x,\alpha}
  U_{x+\alpha,\alpha}\ldots U_{x+(n-1)\alpha,\alpha} \Phi_{x+n
    \alpha})\nonumber \\ 
  1^-&: & V_{x,\alpha}^b (n) =\frac12 Tr
  (\tau^b \Phi^+_x U_{x,\alpha} U_{x+\alpha,\alpha}\ldots
  U_{x+(n-1)\alpha,\alpha} \Phi_{x+n \alpha})  
  \label{eq:operators}
\end{eqnarray}
For the mass fits we have used the maximum $n(=4)$ for the vector
boson which gave the best signal in the symmetric phase. In the Higgs
channel we did not observe such a strong $n$ dependence. In the
analysis we used both $n=0$ and $n=4$.
\subsection{Implementation on QUADRICS parallel computers}

The simulations which are reported in this paper have been performed on
QUADRICS parallel computers at the University of Bielefeld, at the
QUADRICS Q4o and the CRAY-YMP of HLRZ in J\"ulich.
Additionally, codes have been partly tested on the Q1 at DESY--IfH Zeuthen.
The computing facilities at Bielefeld were provided by the Deutsche
Forschungsgemeinschaft (DFG) to the groups being part of the DFG task
force program ''Dynamical Fermions''.
The code written in TAO had to be portable to various topologies of the
QUADRICS family (Q4o, QH2) of parallel machines. The Q4o is a 32 node
machine with a $2 \times 4 \times 4$ topology, the QH2 has 256 nodes in
$4 \times 8 \times 8$ topology.

For the calculation of large Wilson loops we have defined long strings
of links and their mean field values (as auxiliary field attached to
the point of beginning). The factors must be communicated from remote
processor nodes. Forming loops in a later step, these auxiliary
building blocks must be multiplied non--locally where nodes far away
have to be addressed simultaneously.  Due to memory constraints (size
of the auxiliary field) we were able to measure Wilson loops on the
Q4o for lattice sizes up to $40^3$.  Similar communications have to be
considered for correlation functions on non--neighbouring nodes.

The accumulation of histograms of bulk variables as well as Wilson loops
must be done on the host computer (after taking the appropriate global sums
which are then available on all nodes simultaneously) since integer
arithmetics is not possible on the QUADRICS processors.

\subsection{Multihistogram technique and phase separation}

As usual, the search for the phase transition point requires extensive
application of the multihistogram technique \cite{fs,BunkEA} to the
bulk variables listed above. We have processed data obtained in up to
$7$ runs with different $\beta_H$ values (for a given volume) within
and near to the metastable region. Each run consisted out of $30000$
to $145000$ measurements, depending on the measured integrated
autocorrelation time $\tau_{int}$. The respective maximum of the
autocorrelation time increases roughly linearly with the volume. These
numbers, observed in the run nearest to the respective pseudocritical
$\beta_{Hc}$, range from $200$ (for $30^3$) to $1500$ (for $64^3$).
These are autocorrelation times {\it with} tunneling and mainly
indicate the longer lifetime of each one of the two metastable phases,
{\it i.e.} the suppression of tunneling in the larger lattices.

We look for the phase transition point in terms of the critical hopping
parameter, $\beta_{Hc}$, for various (in fact two, in the present work)
values of the lattice gauge coupling $\beta_G$, while the continuum
couplings are kept fixed. In other words, we have studied the phase
transition driven by $m_3$. Then the lattice Higgs self--coupling
$\beta_R$ varies with $\beta_H$ (see (\ref{eq:betar})). Therefore, the
reweighting uses not only ${E_{link}}$, but ${\rho^2}$ and ${\rho^4}$
at the same time. In general, a $3$--dimensional binning has to be
performed in the two relevant parts of the action (corresponding to
$\beta_H$ and $\beta_R$) and some other observable of interest. This
enables to create histograms for any bulk variable and for any value of
$\beta_H$ near to the transition point, based on an estimated density
of states which subsums all measurements in the metastable region with
appropriate weights.

We have determined the finite volume pseudocritical $\beta_{Hc}(L)$ by 
the minima of the Binder cumulants
\begin{equation}
  \label{eq:binder_cum} B_{E_{link}}(L,\beta_H)= 1 - {\frac{{\langle
{E_{link}}^4 \rangle} }{{\ 3 {\langle {E_{link}}^2 \rangle}^2}}}
\end{equation}
and $B_{\rho^2}(L,\beta_H)$, by the maxima of the susceptibilities
\begin{equation}
\label{eq:suscept} C_{E_{link}}(L,\beta_H) = \langle {E_{link}} ^2
\rangle  - \langle {E_{link}} \rangle ^2
\end{equation}
and $C_{\rho^2}(L,\beta_H)$, and using the equal weight method. Our
aim is to apply this method for phase transitions of relatively weak
first order. For these transitions a frequent tunneling between the
pure phases is observed in the metastability region for all lattice
sizes studied. This has made necessary the refinements described
below.

Concerning histogram methods at first order transitions, there is an
ambiguity in the literature how to define the critical coupling (in
fact: pseudocritical for finite volume).  Throughout the metastable
region, a two--state signal is visible in the histograms, {\it e.g.}  of
${\rho^2}$ and ${E_{link}}$, but there are {\it two} prescriptions for
$\beta_{Hc}$: equal height of the maxima vs.  equal weight of the two
competing phases. While theoretical arguments \cite{weight} favour the
equal weight criterion, this requires in practice a procedure to
separate the (measured or reweighted) histogram into contributions
from configurations to be attributed to the respective pure phases. In
addition, there are inhomogeneous (mixed) configurations contributing
to the histograms.  There is no generally accepted procedure that
unambiguously defines the weights of these three contributions.

Our main assumption is that the pure phases can be described by
Gaussian distributions for any volume averaged quantity. Order
parameters like ${\rho^2}$ or ${E_{link}}$ can be considered in this
context. We have utilized the equal weight method as described below
for the case of ${\rho^2}$. The normalized histogram has been
presented as a weighted sum of three histograms referring to the two
pure phases and to all inhomogeneous mixed states
\begin{equation}
p({\rho^2},\beta_H) = w_b p_b({\rho^2},\beta_H) +
w_s p_s({\rho^2},\beta_H)
 +  w_{mix} p_{mix}({\rho^2},\beta_H)
\label{eq:histog}
\end{equation}
with
\begin{equation}
w_b+w_s+w_{mix}=1.
\end{equation}
$w_{b,s}$ denotes the weight of the broken/symmetric phase, $w_{mix}$ is
the corresponding weight of the mixed state, all weights are $\beta_H$
dependent. The pseudocritical $\beta_{Hc}(L)$ is then found for
$w_b=w_s<0.5$.

The positions, widths and weights of the pure phase histograms at a given 
$\beta_H$ have been obtained by fitting the outer flanks of the  two--peak 
histogram to Gaussian shape. At the same time, this fixes the weight 
$w_{mix}$ and the ${\rho^2}$ distribution to be attributed to configurations 
with domains of both phases in equilibrium. We should mention potential 
additional problems to this fit procedure when histograms of both phases are 
strongly overlapping. In asymmetric transitions (generic for the Higgs case), 
with susceptibilities (widths) very different in both phases, this happens 
for smaller lattice sizes. For sufficiently large volumes one can avoid this 
problem.

We have used an iterative procedure to find the critical $\beta_{Hc}$
according to the requirement $w_b = w_s$. It consists of merging all data
into a single histogram at a tentative $\beta_H$. This step is followed by a
fitting procedure as described above which tells the weights of the pure
phases at this $\beta_H$. If they are not equal, $\beta_H$ is corrected
accordingly.

Phase separation of measurements is necessary not only for
this application of the multihistogram method. Another
particular example is the
measurement of pure phase correlation lengths near to the phase transition.
For this purpose and for the splitting of badly separating histograms
in some particular variables it is
better to use the Monte Carlo time sequence of configurations for
runs in the metastability range. The aim is to remove successful tunneling
escapes and unsuccessful tunneling attempts towards the ''wrong'' phase from
what should then be considered as the Monte Carlo trajectory restricted to the
''right'' phase. The procedure rescans the records of the volume averaged
${\rho^2}$ which has a well separated two--peak signal for all
considered volumes. Referring to this variable a lower cut for the upper
(broken) phase and an upper cut for the lower (symmetric) phase can be
chosen. 
These cuts are determined in such a way that the remaining histograms
(for the ''pure'' phases) are almost symmetric around their maxima. 
If the Monte Carlo history of ${\rho^2}$ enters the range
of a certain phase and stays there for more than $100$ iterations, all
measured quantities (recorded for the whole trajectory, including
correlators etc.) are considered to contribute to the statistics of the
given phase until the trajectory leaves this phase again.

The minimal time the trajectory is required to stay within one phase
has to be much larger than the autocorrelation time {\it without}
tunneling but smaller than the autocorrelation time {\it with}
tunneling. The choice of $100$ iterations is consistent with this
requirement.

\section{Localization of the phase transition}

In Tables \ref{tab:runs_critical12} and \ref{tab:runs_critical16} the
statistics is reported for each set of couplings and volumes, that has
been used in the various procedures for the localization of the phase
transition.
\begin{table}[!thb]
\centering
\begin{tabular*}{80mm}{@{\extracolsep{\fill}}|c|r|r|r|} \hline
  $\beta_H$ & $30^3$  & $48^3$        &     $64^3$      \\ \hline
  0.343440  &         &  75000        &                 \\
  0.343480  &  30000  &               &                 \\
  0.343520  &         &  85000        &                 \\
  0.343540  &  50000  &  80000        &                 \\
  0.343544  &         & 120000        &                 \\
  0.343546  &         &               &      90000      \\
  0.343548  &         & 145000        &     120000      \\
  0.343550  &         &               &     100000      \\
  0.343560  &         &  40000        &                 \\
  0.343580  &         & 110000        &                 \\
  0.343600  &  40000  &               &                 \\ \hline
\end{tabular*}
\caption{Data samples used for the localization of
the transition for
$M_H^*=70$ GeV and $\beta_G=12$}
\label{tab:runs_critical12}
\end{table}
\begin{table}[!thb]
\centering
\begin{tabular*}{80mm}{@{\extracolsep{\fill}}|c|r|r|r|}
\hline
  $\beta_H$ & $32^3$  & $40^3$ &  $48^3$      \\ \hline
  0.340700  &         &        &  45000      \\
  0.340780  &  40000  &  40000 &  45000      \\
  0.340800  &  40000  & 100000 &  90000      \\
  0.343820  &  40000  &  40000 &  45000      \\  \hline
 \end{tabular*}
\caption{Data samples used for the localization of
the transition for
$M_H^*=70$ GeV and $\beta_G=16$}
\label{tab:runs_critical16}
\end{table}

\subsection{The infinite volume limit for $\beta_{Hc}$}

In order to determine the set of pseudocritical values of the couplings
$\beta_H$ (and the corresponding $\beta_R$)
we have used the volume averages ${\rho^2}$
and ${E_{link}}$ searching for the minima of the Binder cumulants
(\ref{eq:binder_cum}) $B_{\rho^2}(L,\beta_H)$
and $B_{E_{link}}(L,\beta_H)$, and for the maxima of the susceptibilities
(\ref{eq:suscept}) $C_{\rho^2}(L,\beta_H)$
and $C_{E_{link}}(L,\beta_H)$. In addition to these methods we have used the
equal weight method in the variant described above.

\begin{figure}[!thb]
\centering
\epsfig{file=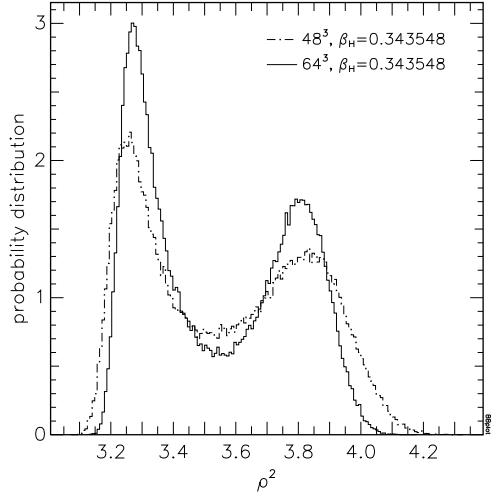,width=6.5cm,angle=0}
\caption{\sl Measured histograms of ${\rho^2}$ for $\beta_G=12$}
\label{fig:rho2hist}
\end{figure}

To demonstrate the two--peak structure we show in Fig.~\ref{fig:rho2hist}
the measured histogram of ${\rho^2}$ on
lattices $48^3$ and $64^3$, all at $\beta_G=12$, for $\beta_H$ values
nearest to the respective pseudocritical $\beta_{Hc}(L)$. The
positions of the maxima change already only slightly with the volume.

\begin{figure}[!thb]
\centering
\epsfig{file=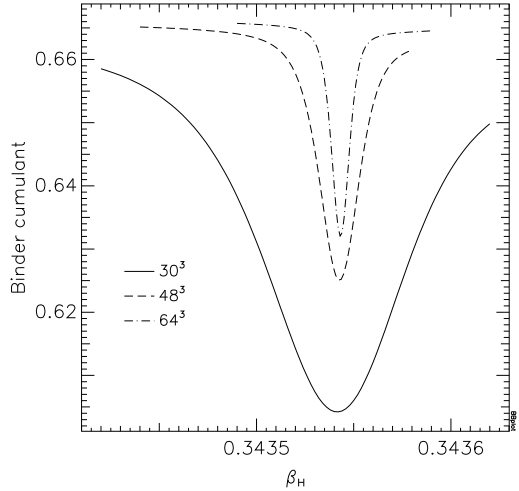,width=6.5cm,angle=90}
\caption{\sl Multihistogram interpolation of $B_{E_{link}}$, $\beta_G=12$}
\label{fig:binder_elink}
\end{figure}

\begin{figure}[!thb]
\centering
\epsfig{file=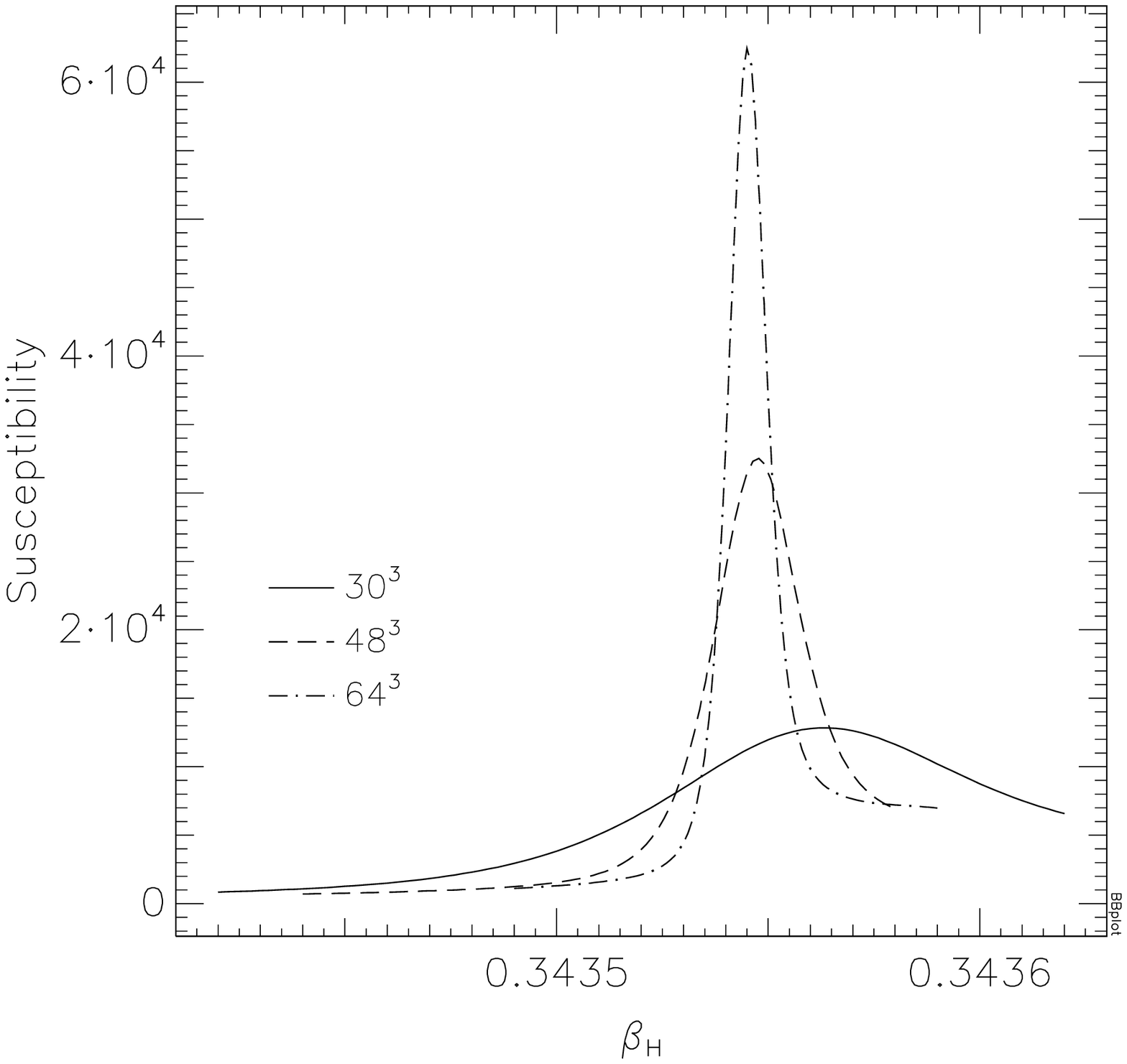,width=6.5cm,angle=90}
\caption{\sl Multihistogram interpolation of $C_{\rho^2}$ , $\beta_G=12$}
\label{fig:susc_rho2}
\end{figure}

In Figs.~\ref{fig:binder_elink} and \ref{fig:susc_rho2}
results of the multihistogram interpolation of our data for
$\beta_G=12$ for the Binder cumulant of ${E_{link}}$ and for the
susceptibility of ${\rho^2}$ are presented. Finiteness and shrinking
of the Binder cumulant with increasing volume present evidence for the
first order nature of the transition at Higgs mass $M_H^*=70$ GeV.
The maximum of the interpolated susceptibility (with the background
susceptibility subtracted) increases almost linearly with the volume.

\begin{figure}[!bth]
\centering
\epsfig{file=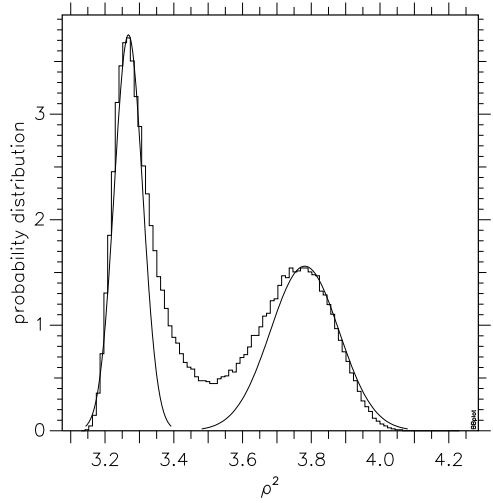,width=6.5cm,angle=0}
\caption{\sl Equal weight histogram of ${\rho^2}$ at pseudocritical
  $\beta_H$, $\beta_G=12$ on a $64^3$ lattice with contributions of the
  pure phases}
\label{fig:hist_rho2_equalweight}
\end{figure}
A histogram reweighted to pseudocritical $\beta_{Hc}(L)=0.3435441$ (as
determined according to the equal weight criterion) for a lattice size
$64^3$ is presented in Fig.~\ref{fig:hist_rho2_equalweight} together
with the Gaussians describing the pure phases just at that
$\beta_{Hc}$. The distribution attributed to mixed configurations with
domains of both phases in equilibrium is well identified between the
two peaks.

\begin{figure}[!thb]
\centering
\epsfig{file=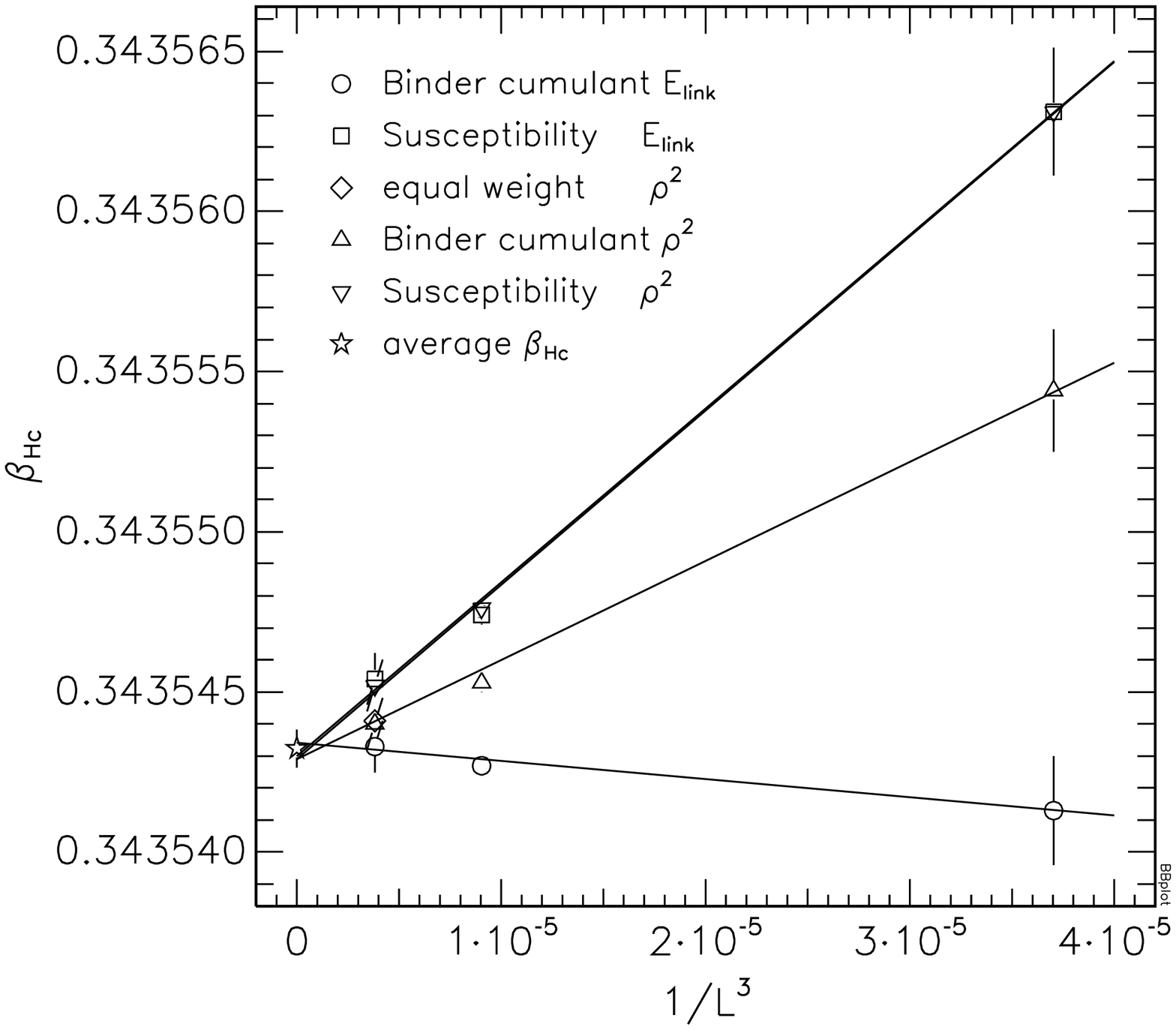,width=6.5cm,angle=90}
\caption{\sl Infinite volume extrapolation of $\beta_{Hc}$ for $\beta_G=12$}
\label{fig:beta_Hc_12}
\end{figure}

\begin{figure}[!thb]
\centering
\epsfig{file=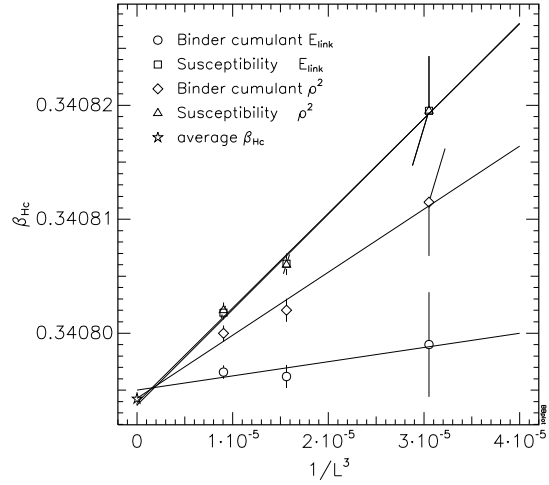,width=6.5cm,angle=90}
\caption{\sl Infinite volume extrapolation of $\beta_{Hc}$ for $\beta_G=16$}
\label{fig:beta_Hc_16}
\end{figure}

The various pseudocritical $\beta_{Hc}(L)$ values for the three
methods applied to ${E_{link}}$ and ${\rho^2}$ are
collected in Figs.~\ref{fig:beta_Hc_12} and \ref{fig:beta_Hc_16}.
They are plotted there versus $1/L^3$ for $\beta_G=12$ and $16$. The
errors of the individual pseudocritical $\beta_H$ values were obtained
by a variant of the Jackknife procedure. From each of the Monte Carlo
histories near to the critical point ($n$ simulated $\beta_H$ values,
$n=3$ in most cases) one half has been left out of the analysis. This
amounts to $2^n$ Jackknife estimators for $\beta_{Hc}$. The equal
weight method could be reasonably applied only for the larger volumes
because of the overlap problem mentioned above.  Corresponding to each
method, a $1/L^3$ fit has been used to yield a respective
$\beta_{Hc}^{\infty}$. The extrapolations nicely coincide as expected.

In table \ref{tab:betah_infty} the extrapolations for each method are
collected together with the average $\beta_{Hc}^{\infty}$ for
$\beta_G=12 $ and $16$. With the use of formulae
(\ref{eq:betah},\ref{eq:laine},\ref{eq:dim_reduct},\ref{eq:4dim_70}) the
$\beta_{Hc}^{\infty}$ is translated into a physical temperature and an
"exact" Higgs mass $M_H$. For definiteness, these numbers are given
for the case of the $SU(2)$ Higgs theory {\it without} fermions.
Comparing these temperatures there seems to be not much space left for
$O(a)$ corrections. The "exact" Higgs mass is practically the same.
\begin{table}[!thb]
\centering
\begin{tabular*}{100mm}{@{\extracolsep{\fill}}|c|rrr|}
\hline
       & $B$ & $C$ & $w_b=w_s$ \\
\hline
\hline
 $\beta_G=12$ & & & \\\hline
$E_{link}$& 0.3435434 & 0.3435430 &               \\
$\rho^2$   & 0.3435429 & 0.3435429 & 0.3435441      \\
\hline
& & & \\
  $\overline {\beta_{Hc}}$ & & 0.3435433(6) &\\
  $T_c/ \mbox{GeV}$        & & 154.34(1) &  \\
  $M_H/ \mbox{GeV}$        & & 66.52     &  \\
\hline
\hline
  $\beta_G=16$ & & &                \\
\hline
$E_{link}$& 0.3407950 & 0.3407937 &             \\
$\rho^2$   & 0.3407943 & 0.3407939 &              \\
\hline
& & & \\
  $\overline {\beta_{Hc}}$ & & 0.3407942(6)  &  \\
  $T_c/ \mbox{GeV}$        & & 154.68(2)   & \\
  $M_H/ \mbox{GeV}$        & & 66.52      &  \\
\hline
\end{tabular*}
\caption{Infinite volume limit for $\beta_{Hc}$ at $M_H^*=70\ \mbox GeV$}
\label{tab:betah_infty}
\end{table}

For comparison, at the smaller coupling ($M_H^*=35$ GeV) the transition
temperature is $T_c=85.7(1)$ GeV with the Higgs mass $M_H=33.0$ GeV.
This has been
obtained for gauge couplings in the range from $\beta_G=12$ to $20$
on lattices of size $40^3$ and $20^3$.

\section{The strength of the phase transition}

\subsection{Condensate discontinuities at the phase transition}

The jumps in $\langle {\rho^2} \rangle$ and $\langle
{\rho^4} \rangle$ are connected to the renormalization group
invariant discontinuities of the quadratic and quartic Higgs
condensates. The two--state signal for $\langle {\rho^2}
\rangle$ and $\langle {\rho^4} \rangle$ is still clearly
visible for all lattice sizes considered at the higher Higgs mass of
$M_H^*=70$ GeV, where the transition turns out much weaker than at
$M_H^*=35$ GeV. The continuum condensate jumps can be put into
relation to the lattice quantities by the following formulae
\begin{eqnarray}
\label{eq:scalar_jumps}
\Delta \langle \phi^+ \phi \rangle/g_3^2 & = & {1 \over 8} \beta_G \beta_{Hc}
\Delta \langle {\rho^2} \rangle \\
\Delta \langle (\phi^+ \phi)^2 \rangle/g_3^4 & = & ({1 \over 8} \beta_G
\beta_{Hc})^2  \Delta \langle {\rho^4} \rangle .
\end{eqnarray}

\begin{figure}[!thb]
\centering
\epsfig{file=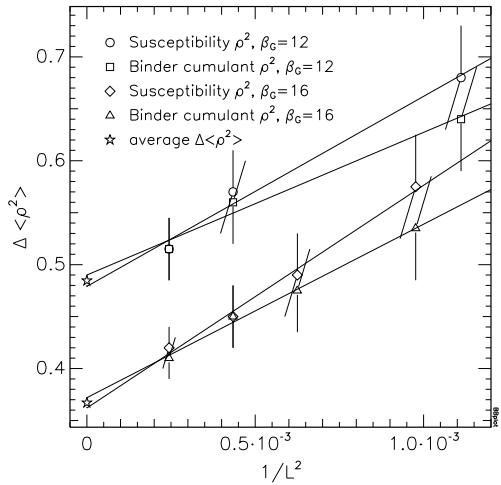,width=6.5cm,angle=90}
\caption{\sl Infinite volume extrapolation of $\Delta \langle
  {\rho^2} \rangle$}
\label{fig:jump_rho2}
\end{figure}

\begin{figure}[!thb]
  \centering \epsfig{file=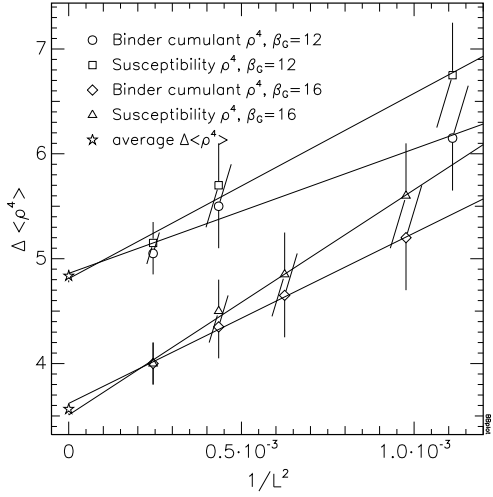,width=6.5cm,angle=0}
  \caption{\sl Infinite volume extrapolation of $\Delta \langle
   {\rho^4} \rangle$}
  \label{fig:jump_rho4}
\end{figure}

Corresponding to the different criteria applied for the definition of
the pseudocritical $\beta_{Hc}(L)$ we obtain histograms of the various
operators just at the respective pseudo--criticality. From the jump of
the operator expectation values between the phases (read off from the
maxima of the corresponding histogram) an infinite volume
extrapolation is performed. A collection of discontinuities of
$\langle {\rho^2} \rangle$ and $\langle {\rho^4}
\rangle$ for various finite lattices is shown in Figs.
\ref{fig:jump_rho2} and \ref{fig:jump_rho4}. As in the analysis of
Ref. \cite {Kajantieoct95} we found that the size dependence of the
jumps for all available lattice sizes is best described by a $1/L^2$
fit. The extrapolations to infinite volume (for the two criteria
$B_{\rho^2}$ and $C_{\rho^2}$) and the averages (of the results
obtained with different
methods to
extrapolate) of the jumps of the
scalar Higgs operators are given in table \ref{tab:jumps_at_70}
\begin{table}[!thb]
\centering
\begin{tabular*}{100mm}{@{\extracolsep{\fill}}|c|rrr|}
\hline
           & $B_{\rho^2}$   &   $C_{\rho^2}$    &   average      \\
\hline
\hline
 $\Delta \langle {\rho^2} \rangle$ & & &\\
\hline
$\beta_G=12$  & 0.490(9)    &    0.479(8)       &    0.485(6)             \\
$\beta_G=16$  & 0.372(8)    &    0.362(7)       &    0.367(5)             \\
\hline
\hline
 $\Delta \langle {\rho^4} \rangle$ & & &\\
\hline
$\beta_G=12$  & 4.86(9)     &    4.81(8)        &    4.84(6)             \\
$\beta_G=16$  & 3.62(8)     &    3.51(7)        &    3.56(5)             \\
\hline
\end{tabular*}
\caption{Infinite volume limit for
  $\Delta \langle {\rho^2} \rangle$ and
$\Delta \langle {\rho^4} \rangle$
at $M_H^*=70$ GeV}
\label{tab:jumps_at_70}
\end{table}
in lattice units.

The corresponding condensate discontinuities in continuum units are collected
in table
\ref{tab:cond_at_70}.
\begin{table}[!thb]
\centering
\begin{tabular*}{80mm}{@{\extracolsep{\fill}}|c|r|r|}
\hline
           & $\Delta \langle \phi^+ \phi \rangle/g_3^2$
           &   $\Delta \langle (\phi^+ \phi)^2 \rangle/g_3^4$
\\
\hline
$\beta_G=12$  &  0.250(3)    &     1.28(2)     \\
$\beta_G=16$  &  0.250(4)    &     1.65(3)     \\
\hline
\end{tabular*}
\caption{The continuum Higgs condensate discontinuities
  at $M_H^*=70$ GeV}
\label{tab:cond_at_70}
\end{table}
The quadratic Higgs condensate is already independent of finite $a$
effects. On the contrary, the quartic condensate shows a severe $a$
dependence. So we conclude, it is more subtle to extract an
appropriate continuum value for this higher condensate.

For the case of the lighter Higgs mass $M_H^*=35$ GeV we present the
corresponding quantities in table \ref{tab:jumps_at_35}.
\begin{table}[!thb]
\centering
\begin{tabular*}{60mm}{@{\extracolsep{\fill}}|c|r|}
\hline
$\Delta \langle {\rho^2} \rangle$ & 6.24(1)\\
$\Delta \langle {\rho^4} \rangle$ & 96.1(2)\\
\hline
$\Delta \langle \phi^+ \phi \rangle/g_3^2$  & 3.20(1) \\
$\Delta \langle (\phi^+ \phi)^2 \rangle/g_3^4$ &  25.2 (1)     \\
\hline
\end{tabular*}
\caption{$\Delta \langle {\rho^2} \rangle$ and
$\Delta \langle {\rho^4} \rangle$ and condensates
at $M_H^*=35 $ GeV
and $\beta_G=12$}
\label{tab:jumps_at_35}
\end{table}
These data were obtained from two separate, completely metastable runs
on a $40^3$ lattice at $\beta_G=12$ and $\beta_H=0.34140$. They failed
to tunnel to the other phase. The actual $\beta_H$ value of these two
runs had been found as pseudocritical one on a smaller lattice.

Besides of jumps in the quadratic and quartic Higgs condensates also a
discontinuity in the expectation values of ${E_{link}}$ appears being
a good indicator for the phase transition as well.

It is known (see \cite{FarakosEA}), that the sum of expectation
values of the various Higgs operators is constant according to a sort
of Schwinger--Dyson equation 
\begin{equation}
- 3 \beta_H \  \langle {E_{link}} \rangle + (1-2\beta_R)
 \langle {\rho^2} \rangle + 2 \beta_R \  \langle 
{\rho^4} \rangle =C
  \label{eq:sd}
\end{equation}
independent of the couplings $\beta_H$, $\beta_G$ and
$\lambda_3/g_3^2$. We have checked this sum rule for all used coupling
values and found $C=2$ exactly within very good numerical accuracy.

From eq. (\ref{eq:sd}) one sees that the Higgs condensate
discontinuities are related to each other by the following sum rule
\begin{equation}
- 3 \beta_H \ \Delta \langle {E_{link}} \rangle + (1-2\beta_R)
\Delta \langle {\rho^2} \rangle + 2 \beta_R \ \Delta \langle
{\rho^4} \rangle = 0 .
\end{equation}

Additionally, the expectation value of the average plaquette $\langle
P \rangle$ shows a discontinuity as well at the phase transition. The
jump in this observable is numerically a tiny effect at the larger
Higgs mass (mostly superimposed by the fluctuations, see the
discussion and Fig.~{\ref{fig:plaq_two_peak} below).  Nevertheless, we
  are able to estimate this jump using the phase separation technique
  discussed earlier.

In table \ref{tab:jump_plaq} the jump $\Delta \langle P \rangle$ is reported
for $\beta_H$ values nearest to the critical ones at $M_H^*=70$ and $35$ GeV
at lattice sizes $64^3$ and $40^3$, respectively.
\begin{table}[!thb]
\centering
\begin{tabular*}{80mm}{@{\extracolsep{\fill}}|rr|r|}
\hline
& & $\Delta \langle P \rangle$ \\
\hline
$M_H^*=70$ GeV & $\beta_G=12$ & 0.00037 \\
  &              $\beta_G=16$ & 0.00015\\
\hline
$M_H^*=35$ GeV & $\beta_G=12$ & 0.00370\\
\hline
\end{tabular*}
\caption{Estimated plaquette jump $\Delta \langle P
\rangle$}
\label{tab:jump_plaq}
\end{table}
At the larger $M_H^*$ the phase separation technique is used.  We do
not set these jumps in correspondence to a continuum gauge condensate
discontinuity
\begin{equation}
 \Delta\langle \frac{1}{4} F_{\alpha \beta}^a F_{\alpha \beta}^a \rangle/g_3^6
= - 3 {\beta_G^4 \over 64}  \Delta \langle P \rangle
\end{equation}
since severe $a$ effects and perturbative contributions are expected.

\subsection{Latent heat}

The latent heat $L_{heat} = \Delta \epsilon$ ($\epsilon$ is the
density of thermal energy) of the transition 
is calculated according to \cite{FarakosEA} 
\begin{eqnarray}
\label{eq:lat_heat}
\frac{L_{heat}}{T_c^4} &=& \frac{M_H^2}{T_c^3} \  \Delta
\langle {\phi^+\phi} \rangle
\end{eqnarray}

With the reported jumps and the values for $M_H$ and $T_c$ ($g^2
(\mu_{T_c})\approx 0.38$) we find the latent heats at the higher and
smaller Higgs masses as given in table \ref{tab:latheat}.
\begin{table}[!thb]
\centering
\begin{tabular*}{60mm}{@{\extracolsep{\fill}}|c|c|}
\hline
& $L_{heat}/T_c^4$ \\
\hline
$M_H^*=70$ GeV &  0.0176(3) \\
$M_H^*=35$ GeV &  0.180(1) \\
\hline
\end{tabular*}
\caption{Latent heat}
\label{tab:latheat}
\end{table}

The equal weight method makes it possible to reconstruct directly the
free energy densities of the pure phases in the vicinity of the phase
equilibrium. The corresponding numbers are obtained in the iterative
search of the pseudocritical $\beta_{Hc}$ using the reweighting
technique at fixed $\beta_G$.  The latent heat can then be expressed
alternatively as the jump $\Delta \epsilon$ of the energy density by
\begin{equation}
{L_{heat} \over T_c^4}={1 \over T_c^2 a^3 L^3} {d \over dT}
\Big(\log w_s
 - \log w_b\Big)\Big|_{T=T_c}.
\end{equation}
Using at fixed $\beta_G$
\begin{equation}
{d \over dT}\Big|_{T=T_c}\approx-\beta_{Hc}^2{a^2 M_H^2 \over 2 T_c}
{\partial \over \partial \beta_H}\Big|_{\beta_H=\beta_{Hc}}
\label{eq:d_over_dT}
\end{equation}
one finds
\begin{equation}
\label{eq:lat_heat_weight}
\frac{L_{heat}}{T_c^4} =
-{g_3^2 \over T_c^3}\ {1 \over 8} M_H^2 \beta_{Hc}^2 \beta_G\
\frac{1}{L^3}{\partial \over \partial \beta_H}
\Big(\log w_s - \log w_b\Big)\Big|_{\beta_H=\beta_{Hc}}.
\end{equation}
The change of the weights very close to the critical $\beta_H$ is
shown in Fig.~\ref{fig:weights_12_64}. Taking the corresponding
slopes needed in (\ref{eq:lat_heat_weight}) we obtain at $M_H^*=70$
GeV for the largest lattices (where the equal weight method can be
applied without ambiguity) $L_{heat}/T_c^4 = 0.0180(9)$ which is in
good agreement with the other method.
\begin{figure}[!thb]
\centering
\epsfig{file=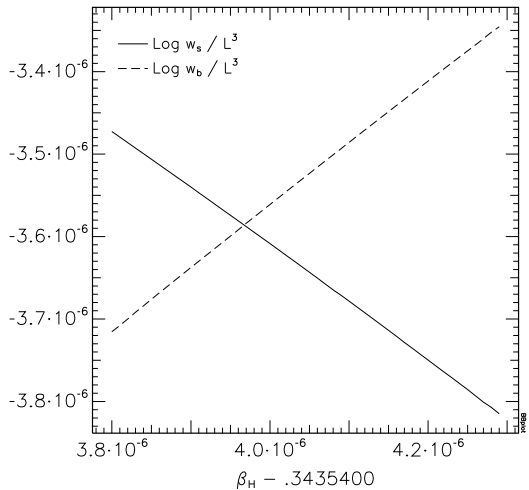,width=6.5cm,angle=90}
\caption{\sl Weights for the pure phases around the equal weight
  pseudocritical $\beta_H$ for $\beta_G=12$ on a $64^3$ lattice}
\label{fig:weights_12_64}
\end{figure}

\subsection{Wilson loops at the phase transition}

If the lattice is large enough, a two--state signal becomes visible
also in the average plaquette $P$.
This can be seen in Fig.~\ref{fig:plaq_two_peak} for $\beta_G=12$,
$\beta_H=0.343548$ at $64^3$. $120.000$ measurements are collected in
this histogram.

\begin{figure}[!thb]
\centering
\epsfig{file=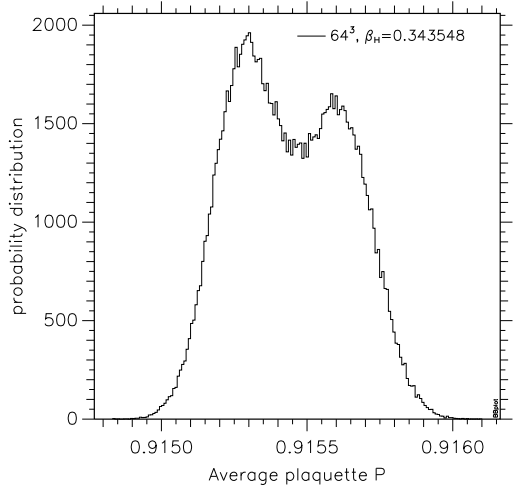,width=6.5cm,angle=90}
\caption{\sl Measured histograms of $P$ for $\beta_G=12$}
\label{fig:plaq_two_peak}
\end{figure}

\begin{figure}[!thb]
\centering
\epsfig{file=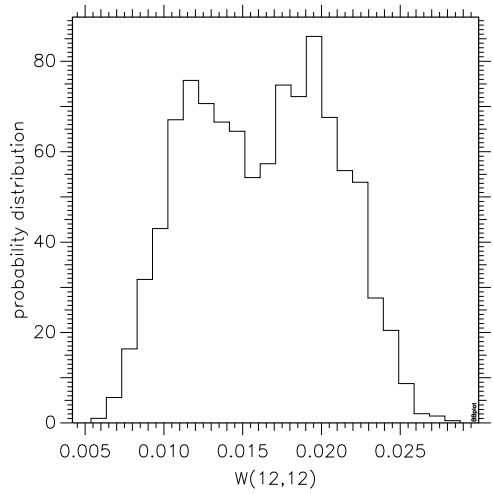,width=6.5cm,angle=0}
\caption{\sl Measured histogram of the $12\times12$ Wilson loop}
\label{fig:wilson_two_peak}
\end{figure}

For medium size rectangular Wilson loops a two--state signal is even
better visible than for the plaquette. This is due to the fact that
the configuration averages follow an area and perimeter law in the
symmetric and broken phase, respectively. If these loops are not too
large ({\it i.e.} not too small numerically), this can be well
separated in the histograms of Wilson loops as demonstrated in
Fig.~\ref{fig:wilson_two_peak}. In that figure a histogram of $2.000$
measurements of a space--averaged Wilson loop of size $12 \times 12$
is shown obtained during $20.000$ MC iterations in the metastability
region at lattice size $48^3$. The corresponding average plaquette for
this size does not show a two--state signal.

\begin{figure}[!thb]
\centering
\epsfig{file=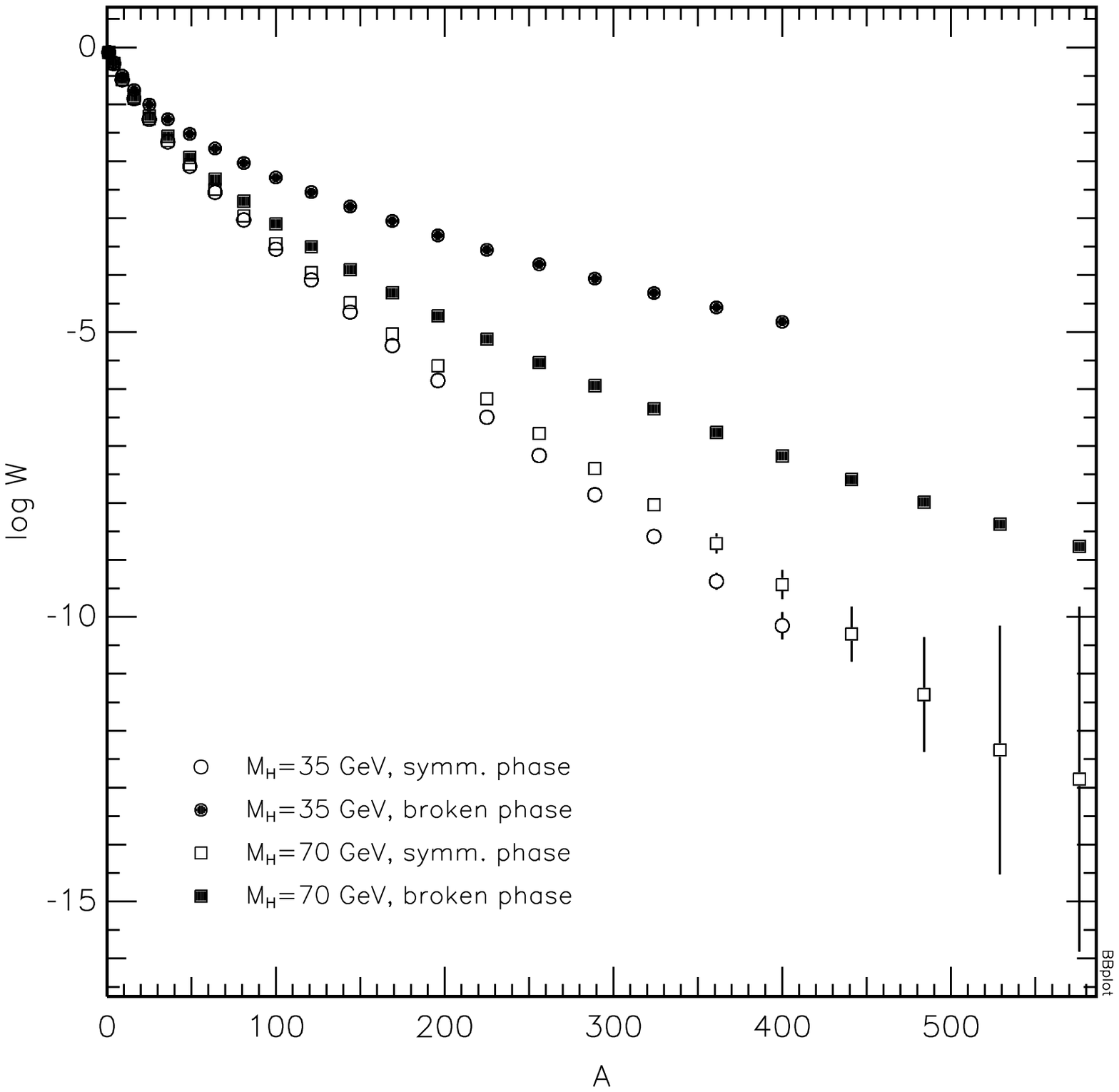,width=6.5cm,angle=90}
\caption{\sl Wilson loops for $\beta_G=12$}
\label{fig:wilson_area}
\end{figure}

In Fig.~\ref{fig:wilson_area} we plot the logarithm of expectation values of
symmetric $L \times L$ Wilson loops versus area $A=L^2$, separated into
symmetric and broken phase contributions at the critical $\beta_H$, for both 
Higgs masses $M_H^*=70$ GeV and $35$ GeV (in both cases for $\beta_G=12$).
The numbers of measurements are $2.000$ in both cases.
This figure shows clearly the area law in the symmetric phase and
practically no (or only weak) dependence  of the string tension (in lattice 
units) on the Higgs self--coupling $\lambda_3$.

\subsection{Estimate of the surface tension}

The coexistence of both phases in lattice configurations opens the
possibility to determine the surface tension $\alpha$.  The use of the
equal weight method as described above allows to estimate the
contribution of the mixed phase state at the pseudocritical coupling
for which $w_b=w_s$ (for large enough lattices). The mixed phase state
weight $w_{mix}$ is directly related to $\alpha$. Using these weights
to estimate $\alpha$ seems to be more natural than to obtain the
surface tension from the ratio of the maximum to the minimum of the
${\rho^2}$ distribution at equilibrium defined by equal weight.  For
asymmetric transitions the choice of the maximum is even somewhat
arbitrary (unless {\it equal height} is the criterium of choice).

To get an idea about the shape of the interfaces (bubbles, walls) we
show a snapshot of a particular configuration on a $32^2 \times 128$
($M_H^*=70$ GeV, $\beta_G=12$) in the pseudocritical $\beta_H$ region.
This configuration has a $\rho^2=3.67$ where  mixed states can be
expected. To suppress the large fluctuations in the local $\rho_x^2$
values we have averaged them iteratively over the next neighbours.
The configuration obtained in this way is shown in
Fig.~\ref{fig:config} where we have plotted all lattice points with
$\rho_x^2>3.67$.
\begin{figure}[!thb]
\centering
\vspace{-4cm}
\epsfig{file=configempty.eps,angle=90,width=12cm}
\vspace{-4cm}
\caption{\sl Smoothed mixed state configuration, plotted are the
  points to be assigned to the broken phase}
\label{fig:config}
\end{figure}
The filled (empty) region can be interpreted as the broken (symmetric)
phase contribution. Besides of a few bubbles we see two large
interfaces separating the pure phases. Although the configuration has
been smoothed the interfaces are rather structured which makes it
difficult to assign them a definite area $A$. For simplicity we use in 
the fit below $A=L_x^2a^2$,
which tends to overestimate $\alpha$.
 
We parametrize the relation between the weights at pseudocriticality
and $\alpha$ for lattices of the form $L_x^2 \times L_z$ as follows
\begin{equation}
\frac{w_{mix}}{w_s}=\frac{w_{mix}}{w_b}= b \; L_z^2 \; \log L_x
\exp(-2\alpha a^2 L_x^2/T_c).
\end{equation}
With that ansatz we try to describe data obtained either from cubic and
from cylindrical lattice geometries.

The factor $L_z^2$ is an entropy factor which accounts for the
positions of the two surfaces, $\log L_x$ is the result (for $d=3$) of
the capillary wave approximation for the (internal) fluctuations of the
surfaces.
The degeneracy factor $b$ is introduced to count different
possible orientations of the surfaces dominant for cubic ($b=3$)
and for prolongated lattices ($b=1$).

In Fig.~\ref{fig:surfacetension}
\begin{figure}[!thb]
\centering
\epsfig{file=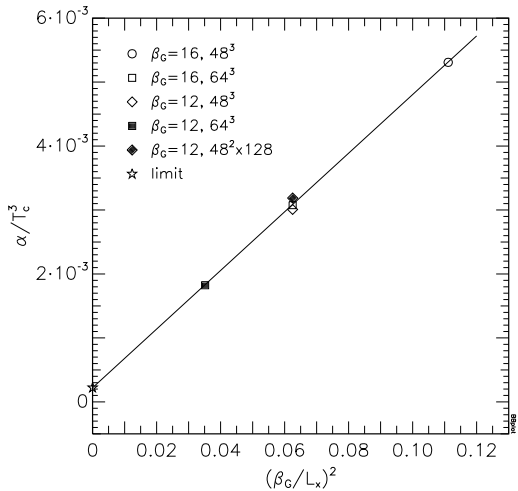,width=6.5cm,angle=90}
\caption{\sl Surface tension vs. $(\beta_G/L_x)^2$}
\label{fig:surfacetension}
\end{figure}
we present the data for $\alpha/T_c^3$ as function of
$(\beta_G/L_x^2)$ for various lattice sizes and geometries and
different $\beta_G$ values at $M_H^* = 70$ GeV. From a linear fit to
the infinite volume limit we find the upper bound
\begin{equation}
\frac{\alpha}{T_c^3} \approx 0.00022.
\end{equation}

These lattice results are smaller by one order of magnitude than the
one--loop estimates for the surface tension \cite{KripfganzEA3}. A trend of
finding smaller surface tensions at larger Higgs masses was already observed in
\cite{Kajantieoct95}. Whereas the latent heat is determined by the position
of the broken minimum alone, the surface tension is sensitive to the shape
of the effective potential in the whole $\varphi$ range between the symmetric
and the broken phase. The disagreement between the measured surface
tension and the one--loop estimate at intermediate Higgs mass appears to
indicate that the loop expansion to the effective potential gets out of
control at intermediate $\varphi$ values already, reflecting the infrared
problems of the symmetric phase.

The surface tension is the only quantity indicating a substantial
deviation from perturbative predictions at $M^*_H=70$ GeV, {\it i.e.}
additional weakening of the phase transition with increasing Higgs
mass. At still larger Higgs masses ($M_H \ge M_W=80$ GeV) the phase
transition has been suggested recently 
to be of second order
\cite{karsch} 
or to be a smooth crossover 
\cite{Kajantiemay96} in accordance with predictions from the study of gap
equations \cite{gap}.
The first case would be difficult to discriminate from a phase
transition being very weakly first order. In any case, however, one will 
observe the turn--over to a very weak or continuous 
transition at higher Higgs
mass.

\section{Broken phase and perturbation theory}

Analysing the lattice data one should answer the question whether the
broken phase can be understood perturbatively. There are no infrared
problems in this phase because the elementary excitations are massive.
Nevertheless, a perturbative treatment is expected to break down at
larger Higgs masses, close to the phase transition.

It is known \cite{KripfganzEA} that for calculating the
temperature dependent effective potential the appropriate effective
expansion parameter for the broken phase is
\begin{equation}
g_{3\,eff}^2=\frac{g_3^2}{2 m_W(T)}
  \label{eq:g3eff}
\end{equation}
with the $3$--dimensional gauge boson mass $ m_W(T)$.  This coupling
can be determined from the two--loop effective potential
\cite{KripfganzEA2}.  Its value depends on the gauge fixing as well as
on the choice of the renormalization point $\mu_3$.  Using Feynman
gauge and $\mu_3=g_3^2$ for $M_H^*=70$ GeV this effective coupling is
found to be $1.10$ at the critical temperature.

In Figs.~\ref{fig:brophasephi2} and \ref{fig:brophasemasses} we
present a comparison of our lattice data with two--loop continuum
predictions for the renormalized Higgs condensate $\phi^2(T)$ and the
vector boson ($m_W(T)$) and Higgs boson ($m_H(T)$) masses at the
corresponding temperature, respectively, using the effective potential
\cite{KripfganzEA2}. The temperature $T$ is implicitly given via
$m_3^2(g_3^2)$ (see eqs. (\ref{eq:dim_reduct}, \ref{eq:4dcouplings}, 
\ref{eq:4dim_70}). The $\langle \phi^2 \rangle$ data are obtained
from the measured $\langle{\rho^2}\rangle$ by subtracting the one--loop
and two--loop counterterms \cite{FarakosEA}
\begin{figure}[!thb]
\centering
\epsfig{file=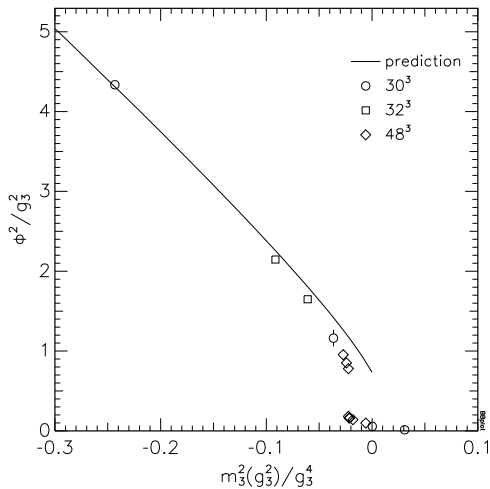,width=6.5cm,angle=0}
\caption{\sl Comparison of perturbative and lattice $\phi^2$}
\label{fig:brophasephi2}
\end{figure}
\begin{figure}[!thb]
\centering
\epsfig{file=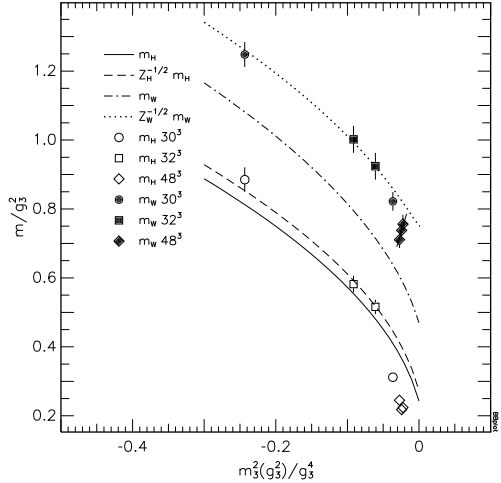,width=6.5cm,angle=90}
\caption{\sl Measured masses at $M_H^*=70$ GeV and $\beta_G=12$
  compared to continuum perturbation theory}
\label{fig:brophasemasses}
\end{figure}
\begin{eqnarray}
\phi^2 & = & 2 \langle \phi^+\phi \rangle(g^2_3), \nonumber\\
\langle \phi^+\phi \rangle(g^2_3)/g^2_3 & = &
\frac{1}{8}\beta_G\;\beta_H \; \langle {\rho^2}\rangle -
\frac{\beta_G}{8\pi} \Sigma - \frac{3}{16\pi^2}
\big(\log\frac{3}{2} \beta_G + 0.67\big).
\end{eqnarray}
We differ from \cite{Kajantieoct95} by a factor of $2$ in the
convention of defining $\phi^2$. The two--loop predictions are
calculated in Feynman gauge.  In the case of $m_W$ and $m_H$
predictions are shown with and without the one--loop wave function
renormalization constants for the Higgs and vector boson masses $Z_H$
and $Z_W$ \cite{KripfganzEA} ($m=Z^{-1/2} m_0$).
The prediction for $m_H$ is derived from the curvature of the effective
two--loop potential at the broken minimum, which is not identical to the pole
mass. However, the difference is expected to be small.

Comparing perturbation theory predictions with the data we observe
very good agreement deeper in the broken phase, as should be expected.
Wave function renormalization is obviously required. Otherwise the
agreement is poor. Close to the phase transition we observe a 
systematic difference between lattice data and the continuum
calculation as function of $m_3^2$.  Nevertheless, the predicted mass
and $\langle \phi^2 \rangle$ values at the corresponding continuum
critical point  agree well with the measured ones at $T_c$
($m_3^2<0$).  The mapping of the two--loop results from the parameter
$v_0 / v$ used in \cite{KripfganzEA2} to $m_3^2(g_3^2)$ fails close to
the phase transition. This may be an indication that higher loop terms
start to play a significant role in this regime.

\section{Some properties of the symmetric phase}

$3$--dimensional $SU(2)$ pure gauge theory is known to possess
confinement, {\it i.e.} the spectrum is formed by massive $W$--balls
instead of the elementary massless $W$'s. Adding a Higgs doublet has
similar consequences as adding fermions: new massive ($\phi^+ \phi$)
bound states occur, and the static confining potential should be
screened. The Lagrangian mass of the scalar bosons increases with
temperature, i.e. sufficiently away from the phase transition the
Higgs bound states become heavy. As a consequence, the symmetric phase
should more and more resemble pure $SU(2)$ gauge theory at higher and
higher temperature.  In the $\beta_H$ range (resp.  $m_3$ range) that
we have explored this could not yet be confirmed, however.  The lowest
Higgs bound state is still significantly lighter than the
lightest $0^{+} W$--ball.

\subsection{Higgs and vector boson bound states}

Results for the lowest Higgs bound state ($0^{+}$) and the vector boson
bound state ($1^{-}$) are shown in Fig.~\ref{fig:H_and_Wmass} as
\begin{figure}[!thb]
\centering
\epsfig{file=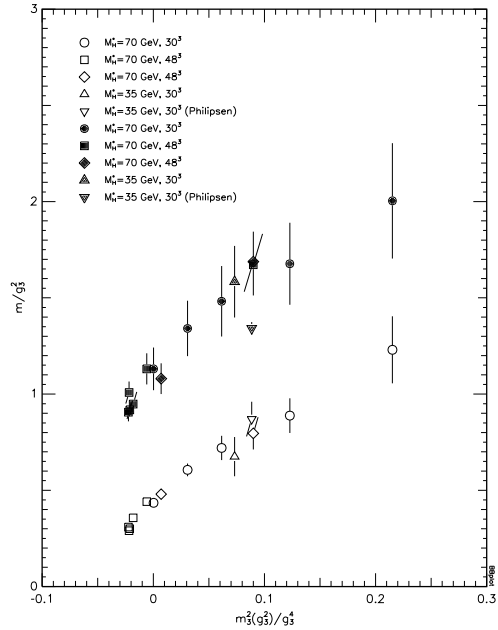,width=6.5cm,angle=0}
\caption{\sl Masses in the symmetric phase for $\beta_G=12$, diamonds
 correspond to $\beta_G=16$}
\label{fig:H_and_Wmass}
\end{figure}
function of $m_3(g_3^2)/g_3^4$. Data from different $\beta_G$ values
nicely coincide. Our results for the vector boson mass $m_W$ should be
considered as upper bounds, because of a possible admixture of states
with a somewhat higher mass. For the Higgs mass $m_H$ this problem is
not severe because the mass gap to higher mass states is significantly
larger. This problem has been carefully studied recently by Philipsen
et al. \cite{Philipsen}.  Note that our gauge boson masses are
calculated from correlators of gauge invariant operators in contrast
to Ref. \cite{karsch} where gauge variant correlators have been
analysed, obtained in conjunction with gauge fixing (Landau gauge).

One important conclusion from Fig.~\ref{fig:H_and_Wmass} is the
scaling behaviour of masses when plotted versus $m_3^2$. Outside the
immediate vicinity of the phase transition no $\lambda_3 $ dependence
is seen within errors. Our data point at the lighter Higgs mass has
been reanalysed as compared to our Ref. \cite{physlett}.

An attempt has been made to reproduce these bound state masses by
approximate analytical methods \cite{Dosch} based on the
Feynman--Schwinger representation of correlators. The measured static
potential (see below) is used as input. The level spacings ($2s$--$1s$,
$1p$--$1s$) are very well reproduced by this approach for the data
point of Ref. \cite{Philipsen}. The other interesting phenomenon is
the $m_3^2$ dependence of the bound state masses, also being
reproduced well.

\subsection{The static potential}

In four dimensions one usually considers Creutz ratios of Wilson loop
expectation values instead of the loops directly in order to avoid
explicit renormalization. This is not necessary in three dimensions
because renormalization becomes very simple. All ultraviolet divergent
contributions are covered by the exponentiated (Abelian) one--loop
term, having a logarithmic divergence. For the potential, the one--loop
massless $W$--exchange contribution is $g_3^2\; {3}/(8 \pi) \log
({R}/(2 a))$. This is the result of a continuum calculation with $a$
introduced as ultraviolet cut--off. Interesting enough, the behaviour
of the two--loop contribution can simply be predicted on dimensional
grounds to be $c_{2l} g_3^4 R$, with some constant $c_{2l}$ not yet
evaluated.  Correspondingly, higher loop orders generate higher powers
of $R$.  Perturbation theory will therefore be appropriate to describe
small Wilson loops, {\it i.e.} the potential at small $R$, whereas it
breaks down in a power--like manner at large distances (infrared
regime). This is different from the $4d$ case where corresponding
effects are logarithmic.

At small $R$, an appropriate fit to the potential (in units of
$g_3^2$) should be provided by
\begin{equation}
\label{eq:potential} V(R) = V_1 + \frac{3}{8 \pi} \log(\frac{R}{2 a}) + c_1 R ,
\end{equation}
where $a$ is identified as the lattice constant, and $V_1$ has been
introduced as additional fit parameter. It could be fixed, however, by
a one--loop lattice calculation (not yet done). The renormalized
potential will anyway contain some arbitrariness in the overall
normalization, due to the choice of the renormalization point.

A priori, it is not clear up to which distance this ansatz may be
appropriate. Interesting enough we shall find (see below) that it
describes our data reasonably well in the whole range, up to about
$R=6/g_3^2$. Two--loop perturbation theory generates a term looking like a
string tension. The value of $c_1$ fitted at intermediate $R$ may
therefore have perturbative as well as non--perturbative contributions.
Because of the gauge coupling being dimensionful it will be difficult
not only in practice but also conceptually to separate perturbative
from non--perturbative effects. This in particular concerns the
definition of a possible $W$ condensate.

Equation (\ref{eq:potential}) is based on massless perturbation theory.
We know, however, that the symmetric phase is formed by massive bound
states instead.  This information may be taken into account in fitting
our potential data. We shall also consider the ansatz ($K_0$ is a
modified Bessel function)
\begin{equation}
\label{eq:potential_K0} V(m,R) = V_2 - \frac{3}{8 \pi} K_0 ( m R ) + c_2 R
\end{equation}
with some additional fit parameter $m$ representing the mass of the
exchanged particle. It should come out close to the measured $W$ mass.

\begin{figure}[!htb]
\centering
\epsfig{file=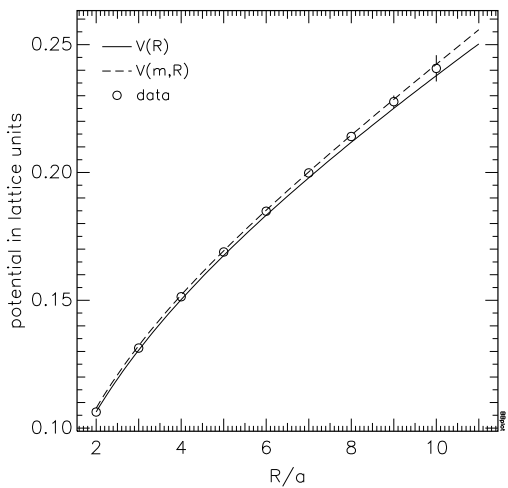,width=6.5cm,angle=90}
\caption{\sl Example for static potential in comparison 
to perturbation theory}
\label{fig:potfit}
\end{figure}
As an example we present in Fig.\ref{fig:potfit} the potential for
$\beta_H=0.3434$, $\beta_G=12$ and $M_H^*=70$ GeV from a run on a
$30^3$ lattice.  The data for $V(R)$ are obtained from exponential
fits to the Wilson loops
\begin{equation}
W( R/a, T/a)= C(R) \exp (-V(R) T) \ , \ \ \  2a \le R \le T - 3a .
\end{equation}
This potential is compared to the above described ans\"atze using
massless ($m=0$) and massive ($m>0$) perturbation theory. For the
second case the mass parameter is chosen to minimize the $\chi^2$
value of the least square fit. For this particular example we obtain
the parameter values $V_1=0.267$, $c_1 = \sigma / g_3^2 = 0.0791 g_3^2$
for the massless and $V_2=0.339$, $c_2= \sigma / g_3^2 =0.118 g_3^2$, 
$m = 0.93 g_3^2$ for the massive case ($m_3^2(g_3^2)/g_3^4=0.00022$). In
general, the fits using massive perturbation theory ansatz seem to
describe the data somewhat better at the larger distances studied.

\subsection{Temperature and Higgs mass dependence of the ``string
  tension''}

In Fig.~\ref{fig:st} we present the string tension obtained from the
ans\"atze for the potential in (\ref{eq:potential}) and
(\ref{eq:potential_K0}), respectively, at different $m_3^2$ values.
\begin{figure}[!htb]
\centering
\epsfig{file=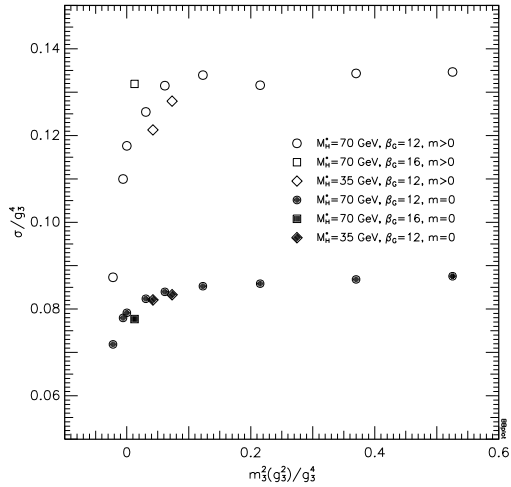,width=6.5cm,angle=90}
\caption{\sl String tension $\sigma/g_3^4$ from (\ref{eq:potential})
  and (\ref{eq:potential_K0}) vs. $m_3^2(g_3^2)/g_3^4$}
\label{fig:st}
\end{figure}
In this analysis data from the higher and smaller Higgs mass $M_H^*$
cases as well as for different $\beta_G$ are used. Most of the data
are from $30^3$ lattices. The closest to $T_c$ and the $\beta_G=16$
string tensions are obtained from fits to Wilson loops on $48^3$ lattices
(compare Fig.~\ref{fig:wilson_area}) up to distances $R=18 a$. The scattering of
the data points is an estimate for the error of the string tension.
The errors obtained from the least square fit are obviously too small
to be reliable.

We observe that there is no significant dependence of $\sigma/g_3^4$
on $\lambda_3$ ($M_H^*$) in both fits.  At larger $m_3^2$
(temperatures), {\it i.e.} deeper in the symmetric phase, the result
of the string tension based on the fit to massive perturbation theory
(\ref{eq:potential_K0}) is (not unexpectedly) close to the value
reported by Teper \cite{Teper} for pure $3d$ $SU(2)$ gauge theory.
This lends additional support to this parametrization of the
potential.
 
For the lattice distances we could explore the expected screening
behaviour of the static potential has not yet been observed. Still
larger distances are difficult to study due to the numerical smallness
of the Wilson loops that are needed.

\section{Summary}

The numerical results of this study provide evidence for the first
order nature of the  thermal phase transition in the $SU(2)$--Higgs
system with Higgs masses up to $70$ GeV.  The answers will soon
converge concerning the upper critical Higgs mass above which the
thermal transition is no longer first order.

One of the motivations for the interest in non--perturbative
investigations of the thermal electroweak phase transition in general,
was to be able to check the reliability of perturbative calculations
of the effective potential. The corresponding results of two--loop
perturbation theory are confirmed for the masses and the renormalized
Higgs condensate in the broken phase. It turns out, that the effect
of wave function renormalization cannot be ignored.  Physics that
depends only on the potential in the vicinity of the broken minimum,
can be systematically improved by higher order perturbation theory.

The surface tension which is sensitive to the barrier shape of the
effective potential for Higgs field below the order parameter at $T_c$
is systematically overestimated in the perturbative calculation
available so far.  But this is not exclusively a problem for
perturbation theory. The surface tension is notoriously hard to
measure for weak transitions as at $M_H \simeq 70$ GeV. The
equilibrium configurations can be inspected and show rather structured
interphase surfaces.  In view of this, estimators of the surface
tension like ours seem to be not extremely justified either.
 
To check the viability of the dimensional reduction program in the
vicinity of the electroweak phase transition has been another motivation
for $3d$ lattice studies. The reduced model does not only make very
precise predictions, but seems to be reliable in the range of Higgs
masses which was of particular interest in this investigation.
In order to learn about the necessity to include higher dimensional 
operators into the effective lattice action one should explore the
case of smaller or very much heavier Higgs masses, but this is much 
less interesting, phenomenologically. In any case, $4d$ anisotropic
lattices will be important for obtaining results (and to study masses)
near to the continuum limit in physically large $3$--volumes.
 
Phenomenologically, the Standard Model is ruled out as an arena for
baryon asymmetry generation at the electroweak transition. The $3d$
lattice approach promises to be applicable as an effective formulation
of nonstandard extensions as well. The phase structure of the model,
once fully reveiled by $3d$ lattice simulations, as well as the
quantitative characterization of the strength of the phase transition,
will be useful to inquire a multitude of $4d$ extended theories by
Monte Carlo parameter exploration.  For this purpose, it is now
possible to go beyond the easiest perturbative formulae.

We have turned the equal weight criterion in conjunction with the
multihistogram Ferrenberg--Swendsen interpolation technique into a
valuable iterative procedure.  In our variant to apply this criterion,
we do not need to choose any {\it a priori} known cut between the
phases nor to associate individual configurations uniquely to one of
the pure phases or to the two--phase mixed configurations. Finally, the
method yields the thermodynamical weight of the latter ones. For any
lattice size where the pure phase histograms do not overlap, the
critical hopping parameter is obtained within the cone of infinite
volume extrapolations of all other, more standard criteria to find 
$\beta_{Hc}$. Moreover, the iterative search for $\beta_{Hc}$ provides 
the thermal energy of both pure phases in the vicinity of the transition, 
allowing to obtain an independent estimator for the latent heat.

We have put much more emphasis than before to the properties of the
symmetric phase. Within the $3d$ approach, information on the spectrum
of particle like states can only be accessed through the $3d$
correlation lengths. The pattern obtained lacks deeper understanding.
Despite some technical progress in resolution of the spectrum of the
$3d$ transfer matrix, the interplay between the confining properties
of the $3d$ effective theory at high physical temperature and the
space--time structure of physical excitations needs further investigations, 
as well as the nature of this confinement itself for
the $3d$ pure gauge 
theory and in the presence of scalar matter fields.  

\section*{Acknowledgements}

E.M.~I., J.~K. and H.~P. are supported by the DFG under grants
Mu932/3-4, We1056/2-3 and Schi422/2-3, respectively. We would like to
thank the system manager M. Plagge of the DFG--Quadrics QH2 parallel
computer for his help. Additionally, we thank the council of HLRZ
J\"ulich for providing CRAY-YMP resources.

\end{document}